# Quasi-periodic pulsations in solar flares: a key diagnostic of energy release on the Sun

Andrew Inglis[1,2], Laura Hayes[3], Silvina Guidoni[1,4], James McLaughlin[5], Valery M. Nakariakov[6], Tom Van Doorsselaere[7], Ernesto Zurbriggen[8], Mariana Cécere[8,9], Marie Dominique[10], Jeff Reep[11], Ivan Zimovets[12], Elena Kupriyanova[13], Dmitrii Kolotkov[6], Bo Li[14], Marina Battaglia[15], Christopher Moore[18], Hannah Collier[15,17], Crisel Suarez[16,18], Tishtrya Mehta[6], Trevor Knuth[1], Thomas Y. Chen[19]

*1. NASA/GSFC, 2. Catholic University of America, 3. ESA, 4. American University, 5. Northumbria University, 6. University of Warwick, 7. KU Leuven, 8. Instituto de Astronomía Teórica y Experimental, 9. Observatorio Astronómico de Córdoba, 10. Royal Observatory of Belgium, 11. Naval Research Laboratory, 12. Space Research Institute of the RAS, 13. Pulkovo Observatory of the RAS, 14. Shandong University, 15. University of Applied Sciences and Arts Northwestern Switzerland, 16. ETH Zürich 17. Vanderbilt University, 18. Center for Astrophysics | Harvard & Smithsonian, 19. Columbia University*

## Synopsis

Solar flares are among the most powerful and disruptive events in our solar system, however the physical mechanism(s) driving and transporting this energetic release are not yet fully understood. An important signature associated with flare energy release is highly variable emission on timescales of sub-seconds to minutes which can often exhibit oscillatory behaviour, features collectively known as *quasi-periodic pulsations (QPPs)*. To fully identify the driving mechanism of QPPs, exploit their potential as a diagnostic tool, and incorporate them into our understanding of solar and stellar flares, new observational capabilities and initiatives are required. There is a clear community need for flare-focused rapid cadence, high resolution multi-wavelength imaging of the Sun, with high enough sensitivity and dynamic range to observe small fluctuations in intensity in the presence of a large overall intensity. Furthermore, multidisciplinary funding and initiatives are required to narrow the gap between numerical models and observations.

QPPs are direct signatures of the physics occurring in flare magnetic reconnection and energy release sites causing periodic behaviour, and hence are critical to understand and include in a unified flare model. To date, despite significant modelling and theoretical work, no single mechanism or model can fully explain the presence of QPPs in flares. Moreover, it is also likely that QPPs fall into different categories that are produced by different mechanisms. At present we do not have sufficient information to observationally distinguish between mechanisms. The motivation to understand QPPs is strengthened by the geo-effectiveness of flares on the Earth's ionosphere, especially if in resonance with geophysical periodicities, and from a solar-stellar perspective by the fact that stellar flares exhibit similar QPP signatures. QPPs present a golden opportunity to better understand flare physics and exploit the solar-stellar analogy, benefiting both astrophysics, heliophysics, and the solar-terrestrial connection.

# Introduction

Solar flares are among the most energetic phenomena in the solar system, releasing up to $10^{32}$ ergs of stored magnetic energy in a matter of minutes to hours. Fully understanding the energy release process and its impacts on the Sun and space environment is a fundamental goal in heliophysics. Despite considerable advances in our understanding of the general picture of solar flares, the detailed properties of the energy release, particle acceleration, and transport processes are not fully understood, and a unified understanding of fundamental flaring processes remains elusive. One key feature that could reveal information about these processes is that the emission associated with flaring energy release is bursty and highly time-dependent on timescales of minutes, seconds or even sub-seconds. In many cases, the emission is modulated with a quasi-oscillatory pattern; such signatures are known as *quasi-periodic pulsations* (QPPs).

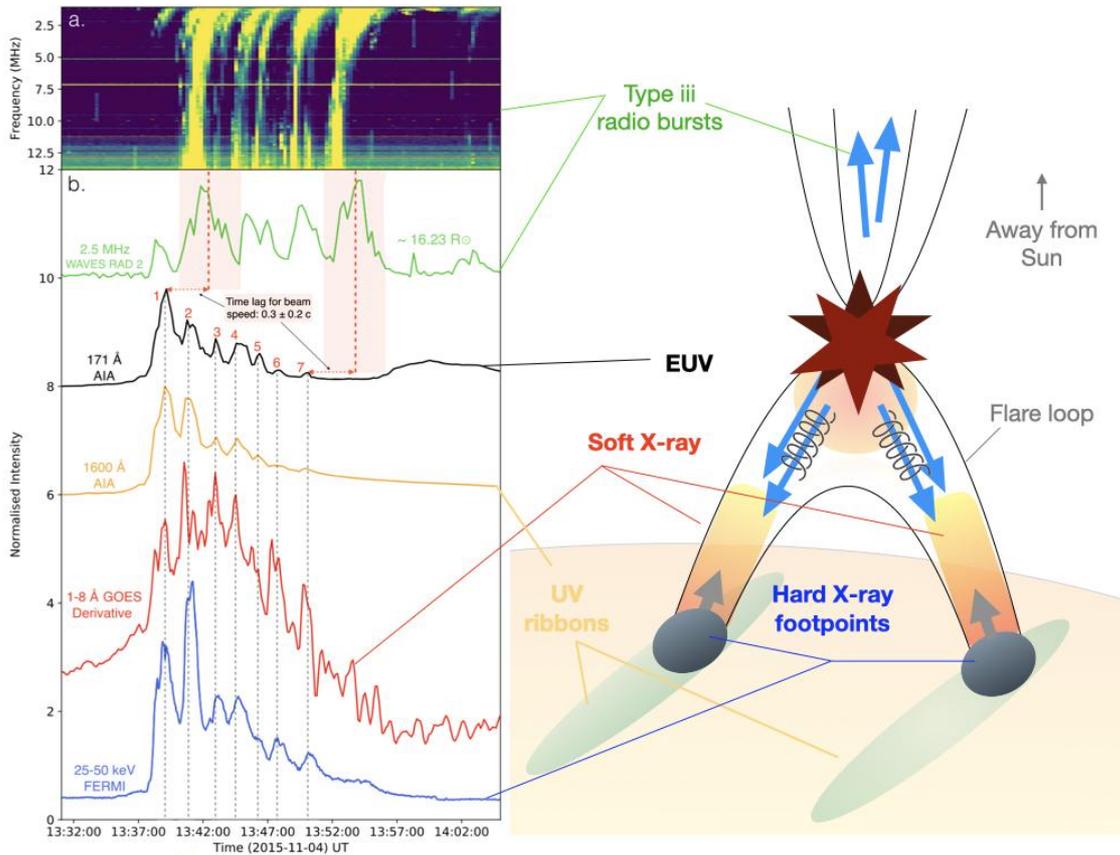

*Figure 1: Left: Quasi-periodic pulsations at a wide range of wavelengths in a solar flare from 04 November 2015 [2]. Right: This cartoon highlights that QPPs in flare emission are associated with the entire flaring process and from different emission mechanisms.*

QPPs are most often identified during the impulsive phase of solar flares in the emission associated with flare-accelerated electrons such as microwave, radio, and hard X-ray observations. However, QPPs co-exist across the entire electromagnetic spectrum, including extreme ultraviolet (EUV) and soft X-ray observations associated with heated coronal and chromospheric plasma, and extending from decimetric radio to even gamma-rays. There is growing evidence that QPPs are an inherent feature of solar flare emission [2, 3], and that they can either manifest as approximately stable or rapidly evolving in time. They have also been identified in both the smallest microflares and the largest X-class flares – with energies in the realm of powerful flares on other magnetically active stars – and occur during the pre-flare phase, the flare impulsive energy release phase, as well as late into the decay phase of flare emission. Some observational studies [4, 5] have also suggested that some QPPs could be related to newly reconnected downward loops after flare eruptions, and also possibly with Supra-Arcade Downflows (SADs). Since the presence of QPPs essentially encompasses all aspects of the flaring process from the energy accumulation and release process to the transport and subsequent heating, their observational characteristics are a direct link to the physical processes involved in solar flares and a valuable source of diagnostic information. The associated timescales and quasi-periodic patterns of QPPs imply the presence of underlying oscillatory or periodic drivers in flares, or repetitive self-organising processes ("self-oscillations"), crucial details that models of flare energy release do not yet – but must – account for. The presence of field-aligned coronal plasma non-uniformities in flare active regions and their essentially "elastic" nature also allows for resonant and dispersive behaviour of various magnetohydrodynamic (MHD) wave processes, a natural mechanism for observed quasi-periodicity. Regardless of the mechanism, no flare model is complete without a full description of the QPP phenomenon.

The presence of QPPs with similar properties in both solar and stellar flares is a key motivator and provides a unique opportunity to investigate the solar-stellar analogy. Solar observations of QPPs, which are temporally well resolved and often observed by multiple instruments, are a valuable asset for interpreting stellar flare QPPs, which are primarily observed in white light with limited cadence, and episodically in other electromagnetic bands [6,7,8]. Well-pronounced QPPs have been identified in stellar flares by multiple authors [9,10,11], with identifications becoming more common due to new observations from surveying instruments such as TESS [12]. Furthermore, the apparent similarity between solar and stellar QPPs [13] suggests common underlying mechanisms. Hence, a better understanding of QPPs in flares from our Sun allows us to leverage the solar-stellar connection and apply our knowledge to stellar coronae active regions, which we cannot directly observe. Depending on the emission mechanism, measuring QPP properties can allow us to infer the properties of stellar plasma such as magnetic field strength and plasma density, properties that cannot be directly observed. From a space weather perspective, it is also now clear that X-ray QPPs during a flare can cause quasi-periodic electron density variations in the Earth's lower ionosphere (D-region), demonstrating their geo-effectiveness [14].

There are a growing number of models that attempt to explain the phenomenon of flare pulsations [15,16], but we are not yet able to conclusively identify the underpinning mechanism driving QPPs. Observational studies of QPPs can be interpreted by several proposed models but we cannot make an unambiguous choice between mechanisms. The challenge is two-fold; observational limitations hinder our ability to perform full temporal, spatial and spectral analysis of QPPs across multiple wavelengths to constrain models, and many models are qualitative rather than quantitative, meaning that we cannot directly compare observations to model outputs, which are typically ran for ideal situations not necessarily close to realistic flaring conditions. The need for coordinated multi-wavelength observations is particularly acute as it allows us to capture thermal and nonthermal plasma emission together - a key requirement for testing models. In this white paper, we illustrate the scientific importance of fully understanding the QPP phenomenon and identify the observational and modelling improvements needed to solve this key science issue.

## Current Science and Limitations

### The observational picture

Current generation solar physics missions allow us to perform time domain analysis of solar flares at cadences of a few seconds. However, to achieve this we mainly rely on full-Sun integrated measurements, such as X-ray data from the GOES/XRS or Fermi/GBM instruments. These instruments do not spatially resolve the Sun, thus using this data we cannot localise where in flare structures pulsations are occurring. For instruments that do image solar flares (such as SDO/AIA), imaging cadences are typically insufficient to capture these phenomena, or lack the imaging dynamic range (Solar Orbiter/STIX) or spatial resolution (radio observations) needed to localise small-amplitude pulsations in the context of a bright flare. Furthermore, images from some instruments (e.g. SDO/AIA) are routinely saturated during even moderate flares (see Fig. 2, right panel), meaning any possibility for localising rapid changes in emission is lost.

Quasi-periodic behaviour is common in flares. Focusing only on the strictly periodic (stationary QPPs), a recent survey of over 5000 solar flares observed in soft X-rays by GOES/XRS during the last solar cycle revealed that over a third of them exhibited evidence of periodic structure in their thermal emission [17]. Such periodicities are consistent with a thermal response to intermittent or periodic magnetic reconnection leading to bursts of particle acceleration. Many events show periods of 10-20s, with some < 10s, approaching the detection limit (see Fig. 2, left). The period distribution *strongly suggests that periodic signatures extend to shorter periods beyond the 4 s detection limit of GOES/XRS*. Furthermore, it is predicted from MHD flare simulations that periodic reconnection may occur on timescales of a few seconds [18], which should produce an observable response in hot flare plasma. The nature and spatial distribution of fast pulsations of thermal plasma can only be uncovered via fast cadence, spatially resolved observations, which are currently unavailable. Other work with Sun-integrated X-ray data has shown that quasi-periodic pulsations are present with even shorter periods of ~1s or even less [19]. Thus, there is a clear, outstanding problem that needs to be solved regarding the dynamics of flare energy release.

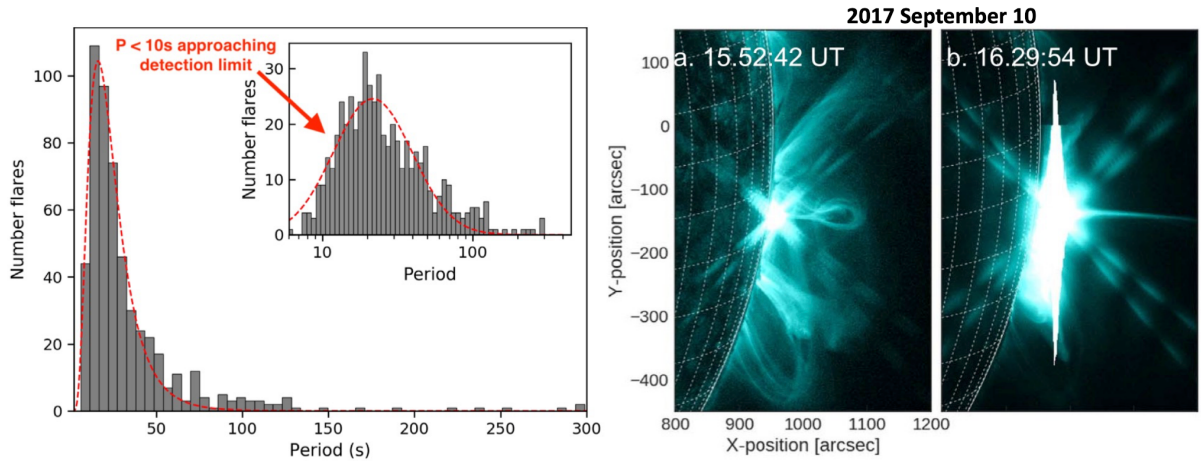

*Fig. 2 - Left: Distribution of periods detected in solar flare thermal emission observed by GOES/XRS between 2011 – 2017 at 2s cadence [17]. Typical periods are 10-20s. Right: SDO/AIA 131A images of the 2017 September 10 X8.2 flare at different times [4]. The EUV data suffers from extreme detector saturation, leaving no possibility of recovering spatial information about observed pulsations in this flare.*

QPPs occur not only in thermal emission but are a common feature in non-thermal emission as well (for example in hard X-rays > 20 keV, and in radio emission). This is a more direct signature of flare energy release and establishes a direct link with energetic particles. Non-thermal and thermal QPPs can often co-exist in the same solar flare. It is unclear whether they are generated by the same or different mechanisms. With this in mind, it is crucial not only to have spatially resolved high cadence observations available; this must be done at multiple wavelengths to capture the thermal and non-thermal regimes simultaneously. Only with simultaneous multi-wavelength observations can we disambiguate the causes of thermal and non-thermal QPP signatures.

## Models and their predictions

From the modelling perspective, the possible drivers of QPPs during flares can generally be categorised into two broad groups: (1) time-dependent regimes of magnetic reconnection and energy release (see Figure 3), and (2) magnetohydrodynamic (MHD) oscillations in flare structures. These broad categories can be further subdivided into more than a dozen specific mechanisms (see [15, 16]).

At present, most of the models that produce QPP signatures are of a qualitative nature – it is not yet possible to quantitatively obtain all the necessary observational properties of the models for realistic conditions in flare regions. A serious drawback of most models is that they are developed within the MHD approximation and do not take into account the particle acceleration process, which plays a very important role in flares. The path forward involves developing realistic 3D models, considering inhomogeneities along the flare polarity inversion line (PIL), the processes of acceleration and propagation of particles, as well as direct forward modelling of the emission of flare regions in different spectral ranges for a detailed comparison with observations of various instruments. Another key issue – as discussed above – is that modern observations usually do not yet provide all the necessary information about the physical properties of flare regions and sources of QPP. It is extremely important to have detailed information about the spatial structure of QPP sources and their dynamics in different spectral ranges as input for the next generation of models. Reliable information on the geometry and dynamics of magnetic fields is also important. Another difficulty relates to the existence of a broad variety of QPP types (classes) with distinctly different observational properties, the identification of which is crucial for direct comparison with theory and revealing the underlying QPP models [20].

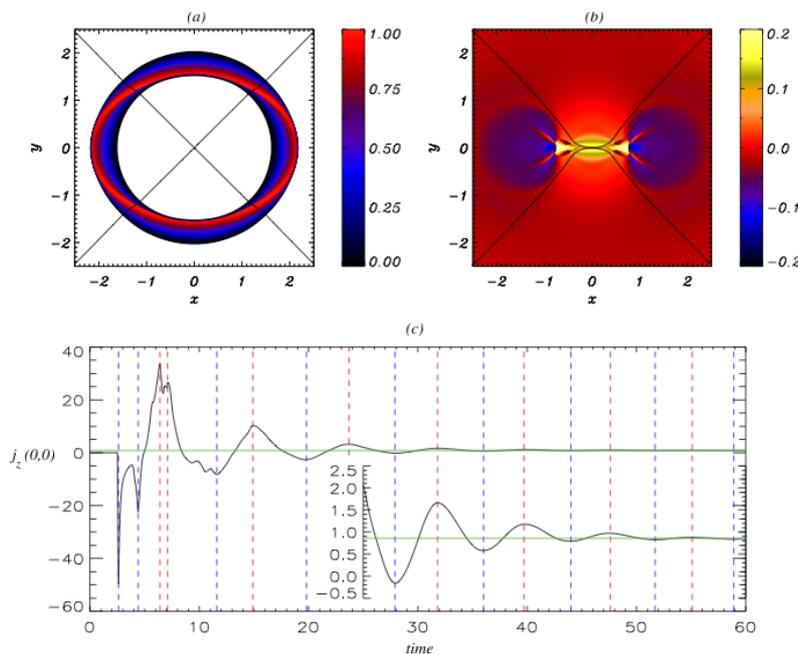

Figure 3: An MHD simulation of periodic reconnection at a magnetic X-point. Top: Contours of perpendicular velocity at two different times. Bottom: Current density $j_z$ at the center of the simulation. $j_z$ is modulated with a period of ~8s, displaying a bursty, damped time profile, qualitatively similar to X-ray and radio observations of flares [21].

Refining models to fully explain flare pulsations requires creating a positive feedback loop between models and observations, so models can produce actionable predictions, and new observations can provide more direct feedback to models. A further long-term goal is to move towards a unified modelling system, where models in different regimes can be chained together to describe the full flare picture.

## Open Questions

Ultimately, we want to answer the question: 'What physical processes produce QPPs in flares?' Doing so will directly improve our understanding of energy release, both on the Sun and in other stars. To achieve this there are several intermediate questions regarding solar flare QPPs that should be answered in the next decade. These are:

1. Where are the sources of QPPs located in flare structures? How is this different between the thermal and nonthermal emission regimes?
2. Are QPPs detected in different phases of the flare fundamentally different? Are QPPs detected in thermal and non-thermal emission different?
3. How can we test and verify numerical models of QPPs, and determine which models are most appropriate?
4. Are QPPs detected in stellar flares – which are much more powerful than the strongest detected solar flares – produced by the same mechanisms as in solar flares?

| Open Questions | Current Limitations | Future needs |
| --- | --- | --- |
| Q1. Where are the sources of QPPs located? | • Insufficient X-ray dynamic range and spatial resolution to image faint sources and QPPs<br>• Image saturation and pixel bleeding in multiple wavelengths during flares<br>• Insufficient imaging cadence at EUV/UV wavelengths | • Focusing X-ray imaging with dynamic range >100:1 with at least 1s cadence<br>• Flare-focused short exposure-time imaging with limited saturation<br>• Rapid cadence EUV imagers (~1s or less)<br>• Spatial resolution of ~ 1 arcsecond |
| Q2. Are QPPs in different flares phases fundamentally different? | • Same as Q1<br>• Cadence of X-ray imaging insufficient to identify shorter periods during flare impulsive phase<br>• Limited imaging spectroscopy capabilities at multiple wavelengths | • Same as Q1<br>• High cadence (<0.5s) X-ray observations with high sensitivity<br>• Multi-wavelength imaging spectroscopy in EUV and X-ray (thermal and non-thermal) at 1s or less |
| Q3. How can we test and verify numerical models of QPPs? | • Models are qualitative in nature - difficult to directly test against observables<br>• Insufficient observational data to constrain model inputs (see Q1, Q2)<br>• Model scales are either macro or micro - difficult to capture all physics in one model | • Coordinated efforts between observational and theoretical communities<br>• Funding to develop models to yield actionable predictions for comparison<br>• Enable chained models where one model output feeds another model input |
| Q4. Are QPPs detected in stellar flares produced by same mechanism? | • Limited funding vehicles for solar-stellar science<br>• Limited white light solar flare observations for direct stellar comparison<br>• Limited simultaneous X-ray and white light stellar flare data | • Dedicated interdisciplinary funding opportunities to exploit the solar-stellar connection<br>• Increased white light solar flare observations coordinated with other wavelengths (X-ray) |

*Table 1: Summary of current limitations and future needs to fully understand pulsations in flare energy release.*

The resolution of Questions 1 and 2 can only be achieved with multi-wavelength spatially resolved observations with a high enough cadence to observe pulsations, and sufficient dynamic range and exposure control to avoid saturation from high fluxes. To achieve this in the X-ray regime, a direct, focusing optics imager is needed [22, 23] to provide sufficient dynamic range. The required temporal resolution is of order ~1s, to reliably capture pulsations with periods of 10 - 20s or less. At EUV wavelengths, telescopes resistant to saturation are an additional requirement. To solve Question 3 requires a long-term effort in advancing numerical models to the quantitative stage. Once models achieve this stage they will produce outputs directly comparable to observations, which combined with more advanced observational data will allow us to determine which models explain the QPP phenomenon. To address Question 4, we need to invest in cross-disciplinary solar-stellar science and encourage collaborations in this domain. The current challenges and future needs of the community are summarised in Table 1.

## Summary and Recommendations

Fully understanding the energy release process and its impacts on the Sun and space environment is a fundamental goal in heliophysics. The presence of underlying oscillatory or periodic drivers in flares and various flare-driven perturbations of coronal active regions, which we collectively refer to as quasi-periodic pulsations (QPPs), is a crucial detail that models of flare energy release do not yet – but must – account for. To address this issue, we make the following recommendations:

1. There is a clear community need for rapid cadence, high resolution imaging of solar flares, with high sensitivity and dynamic range to observe small fluctuations in intensity while handling the large total fluxes of flares. This is required in order to localise the sources of pulsations in flare systems. This is the most effective and potentially the only way to disambiguate models of quasi-periodic behaviour in flare energy release.
2. The above capability is needed in both the thermal and non-thermal emission regimes, which requires multi-wavelength observations. This is critical because the thermal and non-thermal emission regimes capture different flare physics. An example combination is joint EUV, soft X-ray, and hard X-ray imaging.
3. The gap between global flare models that incorporate QPPs and observations must be narrowed. Current modelling is sophisticated but is often not designed to directly predict observables. The community should support efforts to better bridge models and observations, such that models can produce actionable predictions, and so that new observations can provide more direct feedback to models. These efforts could be analogous to those undertaken in the space weather community, for example the NASA Space Weather Science Research to Operations to Research solicitation.
4. Exploit the solar-stellar analogy. Research that engages in cross-disciplinary solar-stellar science should be encouraged; this could be achieved via a dedicated solar-stellar science solicitation. Increased white light solar flare observations coordinated with other wavelengths (e.g. X-ray) would enable more direct comparisons with other stars.

Implementing these recommendations would advance fundamental heliophysics knowledge, of which flares are a key component. The insights into flare energy release that will be gained by these recommendations can also be directly applied to stellar flare physics, advancing the goals of the astrophysics community [24]. In this way, we will solve one of the most enduring mysteries of flare energy release.

# Acknowledgements

This white paper was supported by the International Space Science Institute (ISSI) in Bern, through ISSI International Team project #527, "Bridging New X-ray Observations and Advanced Models of Flare Variability: A Key to Understanding the Fundamentals of Flare Energy Release."

# References

[1] Clarke et al. 2021, ApJ, 910, 123, DOI: 10.3847/1538-4357/abe463
[2] Simoes et al., 2015, SolPhys, 290, 3625, 10.1007/s11207-015-0691-2
[3] Dominique et al. 2018, SolPhys, 293, 61: 10.1007/s11207-018-1281-x
[4] Hayes et al. 2019, *ApJ*, 875, 33, DOI: 10.3847/1538-4357/ab0ca3
[5] Yu et al. 2020, *ApJ*, 900, 17, DOI: 10.3847/1538-4357/aba8a6
[6] Doyle et al. 2018, MNRAS, 475, 2842: 10.1093/mnras/sty032
[7] Kolotkov et al. 2021, ApJ, 923L, 33: 10.3847/2041-8213/ac432e
[8] Broomhall et al. 2019, A&A, 629A, 147: 10.1051/0004-6361/201935653
[9] Anfinogentov, S. et al. 2013:ApJ, 773, 156A: 10.1088/0004-637X/773/2/156
[10] Pugh, C. et al. 2016, MNRAS, 459, 3659: 10.1093/mnras/stw850
[11] Jackman et al. 2019, MNRAS, 482, 5553: 10.1093/mnras/sty3036
[12] Ramsay et al. 2021, SolPhys,296, 162: 10.1007/s11207-021-01899-x
[13] Cho et al. 2016, ApJ, 830, 110: 10.3847/0004-637X/830/2/110
[14] Hayes et al. 2017, JGRA, 122, 9841: 10.1002/2017JA024647
[15] McLaughlin et al. 2018, SSRv, 214, 45: 10.1007/s11214-018-0478-5
[16] Zimovets et al. 2021, SSRv, 217, 66: 10.1007/s11214-021-00840-9
[17] Hayes et al. 2020, ApJ, 895, 50: 10.3847/1538-4357/ab8d40
[18] Guidoni et al. 2016, ApJ, 820, 60: 10.3847/0004-637X/820/1/60
[19] Knuth et al. 2020, ApJ, 903, 63: 10.3847/1538-4357/abb779
[20] Nakariakov et al. 2019, *PPCF*, 61, a4024N: 10.1088/1361-6587/aad97c
[21] McLaughlin et al. 2009, A&A, 493, 227: 10.1051/0004-6361:200810465
[22] Christe et al. 2017, NGSPM white paper: arXiv:1701.00792
[23] Shih et al. 2021, AGU Fall Meeting 2020, abstract #SH048-0012
[24] Vievering et al. 2022, Heliophysics Decadal Survey white paper